# Quasi-perfect spatiotemporal optical vortex with suppressed mode degradation

Shunlin Huang,[1, *] Xiong Shen,[1] Renjing Chen,[2] Jun Liu,[1, 2, *] and Ruxin Li,[1, 2]

[1]*Zhangjiang Laboratory, Shanghai 201210, China*
[2]*State Key Laboratory of High Field Laser Physics, Shanghai Institute of Optics and Fine Mechanics, Chinese Academy of Sciences, Shanghai 201800, China*
*Corresponding authors: huangshunlin@126.com; jliu@siom.ac.cn

**Abstract**: Spatiotemporal optical vortex (STOV) carrying transverse orbital angular momentum (OAM) enriches the family of vortex beams and exhibit unique properties. Typically, a high-order STOV with an intensity null degrades into multiple first-order STOVs embedded within a single wave packet during propagation, a phenomenon known as time diffraction or mode degradation. However, this degradation limits the applicability of STOVs in specialized fields. Therefore, the generation of mode degradation-suppressed STOVs (MDS-STOVs) is of significant for both practical applications and theoretical studies. Herein, we theoretically analyze the generation of MDS-STOVs by utilizing a conical phase to localize the energy of the STOV into a ring-shaped structure. For MDS-STOVs with large topological charges (TCs), the ring-shaped profile can be well-maintained, and the rapid expansion of the beam size with increasing TC is significantly suppressed compared to conventional STOVs. As a result, these MDS-STOVs can be regarded as quasi-perfect STOVs (QPSTOVs). Furthermore, QPSTOVs exhibit strong resistance to group delay dispersion (GDD), eliminating the need for precise dispersion control and facilitating their generation and application. This work advances our understanding of the physical properties of light carrying transverse OAM and opens up exciting avenues for the application of STOVs in diverse fields, such as optical communication and quantum information processing.

## 1. Introduction

Vortex light beam with a spiral phase in the spatial domain and an intensity null in the beam profile has attracted significant attention and has been extensively studied. It has been discovered that spatial vortex light beams can carry orbital angular momentum (OAM) [1], with their OAM orientation being parallel or antiparallel to the beam propagation direction, classified as longitudinal OAM. Vortex light beams with longitudinal OAM have found widespread applications in various fields, such as microparticle manipulation [2-4], optical communications [5], microscopy [6].

In recent years, a novel member of the vortex beam family, known as the spatiotemporal optical vortex (STOV) [7-9], has been experimentally demonstrated, although it was theoretically analyzed a decade earlier [10-12]. STOVs exhibit a phase singularity and an intensity null in the space-time domain. Due to their spiral phase in the space-time domain, the energy of STOVs couples between the spatial and temporal domains during propagation. Unlike conventional spatial vortices, STOVs possess intrinsic transverse OAM, with an orientation orthogonal to the beam's propagation direction. This distinctive feature renders STOVs both fascinating and fundamentally different from traditional spatial vortex beams.

STOVs have garnered significant attention and are rapidly advancing [13]. Recently, Endeavors have been made to explore the fundamental properties of the STOVs. For instance, studies have investigated the conservation of transverse OAM in nonlinear processes [14-16], spin-orbit interactions and angular momenta of STOVs [17-20], multilobe structures of diffracted STOVs where one gap corresponds to a topological charge [21, 22], and the time delays and transverse shifts of STOVs during refraction and reflection at planar interfaces [23]. Additionally, the degradation of high-order STOVs into multiple first-order STOVs during propagation has been extensively studied [8, 9, 24, 25], a phenomenon also referred to as time diffraction [11, 18].

Furthermore, significant efforts have been devoted to developing new methods for generating STOVs, which are currently primarily produced using $4f$ pulse shapers. Alternative approaches include photonic crystal slab [26, 27] and slanted nanograting [28]. Moreover, the STOVs are endowed with more degrees of freedom to explore the basic properties of light and to expand their potential applications. Various STOVs with unique characteristics have been proposed, such as vector STOVs with cylindrical polarized state [29], toroidal STOVs with ring-shaped profile [30], and STOV strings comprising multiple STOVs embedded within a single wave packet [22]. Furthermore, to obtain STOVs with non-spreading [31] and time diffraction-free properties [32], spatiotemporal Bessel vortices have been introduced, offering new possibilities for controlling and manipulating STOVs.

The primary characteristic of mode degradation is that a high-order STOV with a topological charge $l$ will split into $l$ first-order STOVs within a single wavepacket. This splitting effect arises from the imbalance between spatial diffraction and dispersion experienced by the STOV pulse during propagation, a phenomenon known as spatiotemporal astigmatism [9, 14]. It is also referred to as time diffraction [11, 18]. For conventional STOVs, mode degradation is an intrinsic property that cannot be avoided, which may restrict their practical applications. On the other hand, it suggests that suppressing the mode degradation effect might be achievable through precise control of the mismatch between dispersion and spatial diffraction. However, such compensation requires accurate measurement and control of both diffraction and dispersion.

Although STOVs are well confined within the inner ring of the wavepacket in non-spreading Bessel STOVs, where the energy is strongest, the mode degradation effect remains evident and is not suppressed. For time diffraction-free spatiotemporal Bessel vortices, the suppression of mode degradation is only effective in the near field, and precise dispersion control is required, which complicates the generation process. Additionally, the mode degradation effect becomes more pronounced as the topological charge decreases. It is important to note that different orders of time diffraction-free spatiotemporal Bessel vortices may require distinct dispersion compensation strategies. Furthermore, the wavepackets of Bessel STOVs exhibit a multi-ring structure, meaning the energy is distributed across all rings. This energy dispersion may limit their practical applications in scenarios requiring high intensity, such as particle acceleration [33].

Herein, we theoretically analyze the generation and propagation of MDS-STOVs using a combination of conical and spiral phases. These MDS-STOVs exhibit a ring-shape structure and remain free from mode degradation during propagation. We also find that they are insensitive to group delay dispersion (GDD), which facilitates their practical applications. Furthermore, the radius of the ring of the MDS-STOV increases moderately with the TC, in contrast to conventional STOVs, where the expansion is more pronounced. In this regard, MDS-STOVs can be considered as quasi-perfect STOVs (QPSTOVs), analogous to the concept of spatial perfect vortices [34]. The generation of MDS-STOVs, or QPSTOVs, is significant for both theoretical studies and practical applications of STOVs.

## 2. Theory and principle

It is well-established that a spherical lens can perform a two-dimensional (2D) Fourier transform (FT) in the spatial domain, effectively serving as a high-speed mathematical spatial FT calculator. This capability establishes a profound connection between mathematics and optics within the spatial domain. Intriguingly, studies have demonstrated that a 2D 4*f* pulse shaper equipped with a focusing lens can achieve a 2D Fourier transform in the spatiotemporal domain [8-10]. This device functions similarly to the spherical lens, thereby bridging the gap between mathematics and optics in the spatiotemporal domain. The 2D 4*f* pulse shaper with a focusing lens can thus be regarded as a mathematic spatiotemporal FT (STFT) calculator. This setup facilitates the generation of on-demand spatiotemporal wavepackets using their corresponding STFT pairs. In this section, we demonstrate that the 2D STFT of a Bessel beam in the *y-ω* domain results in a ring-shaped beam in the *y-t* domain, which constitutes a perfect STOV (PSTOV). An ideal Bessel beam in the *y-ω* domain with unit amplitude in $z = 0$ can be expressed by

$$E(\rho,\varphi) = J_l(r_0\rho)\exp(il\varphi), \qquad (1)$$

where $\rho = \sqrt{y^2 + \Omega^2}$, $\Omega = \omega - \omega_0$, $\omega_0$ is the central frequency, $\varphi = \tan^{-1}(y/\Omega)$, $J_l$ is the *l*-th order Bessel function of the first kind. The 2D STFT of Eq. (1) transform the field $E(\rho,\varphi)$ in the *y-ω* domain into the field $E(r,\theta)$ in the $k_y$-*t* domain, which can be written as

$$E(r,\theta) = FT\{E(\rho,\varphi)\} = \int_0^\infty \int_0^{2\pi} E(\rho,\varphi)\exp[-i2\pi r\rho(\cos(\varphi-\theta))]\rho d\rho d\varphi, \qquad (2)$$

where $r = \sqrt{k_y^2 + t^2}$, $\theta = \tan^{-1}(k_y/t)$, and $k_y$ is related to $y$ by $k_y = y/(\lambda f)$ in the focal plane of a lens with focal length of *f*, *λ* is the wavelength. Substituting Eq. (1) into Eq. (2), and after calculation, the analytical solution of Eq. (2) can be derived as follows [35]

$$E(r,\theta) = FT\{E(\rho,\varphi)\} \propto \frac{i^{l-1}}{r_0}\delta(r-r_0)e^{il\theta}. \qquad (3)$$

From Eq. (3), we can see that the 2D STFT of an ideal Bessel beam in the space-frequency (*y-ω*) domain yields a ring-shaped beam in the space-time (*y-t*) domain. This ring-shaped beam has a radius of $r_0$ and features a vortex with a TC of *l*. Notably, such a ring-shaped STOV exhibits only one intensity null within the wavepacket. Similar to the concept of a spatial perfect vortex [34], this ring-shaped STOV can be considered as a perfect STOV (PSTOV). Note that, generating an ideal Bessel beam in either the $y$-$\omega$ or $y$-$t$ domain is challenging in practice. As a result, a Bessel-Gauss beam is often employed, where a Gaussian profile is superimposed on the Bessel profile to confine the energy of the Bessel beam effectively.

## 3. Results

In the simulation, QPSTOVs are generated using a 4$f$ pulse shaper, a setup commonly employed in laboratories for STOV generation. The schematic of the generation setup is illustrated in Fig. 1. The 4$f$ pulse shaper consists of a grating (1200 rules/mm), a cylindrical lens ($f$ = 300 mm), and a 2D spatial light modulator (SLM) positioned at the Fourier plane of the pulse shaper, serving as a phase mask. The distances between these three components are identical, each set to 300 mm. The modulated pulses, after reflection, are focused by a spherical lens, and the QPSTOVs are obtained at the focal plane

To generate the QPSTOV, a combination phase $ar + l\varphi$, comprising a conical phase $ar$ and a spiral phase $l\varphi$, is loaded onto the 2D SLM. Here, $a$ represents the axicon parameter, and $l$ denotes the topological charge. The detailed calculation process follows a methodology consistent with our previous works [21, 22].

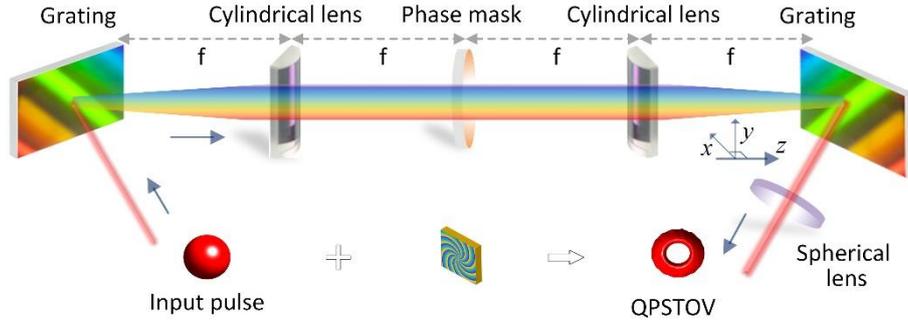

Fig. 1. The Schematic of the experimental setup for the generation of QPSTOVs. The QPSTOVs are generated using a 4$f$ pulse shaper, where a phase mask is positioned at the Fourier plane and combined with a spherical lens to achieve the desired spatiotemporal modulation.

As indicated by Eq. (3), the 2D STFT of a spatial-spectral Bessel beam yields a PSTOV in the space-time domain, with the radius $r$ and the thinness of the ring independent of topological charge $l$. However, our results reveal that the actual space-time wavepacket generated using the 4$f$ pulse shaper does not conform to an ideal PSTOV. Although the radius $r$ and the ring thinness appear nearly unchanged for lower values of $l$ (e.g., $l < 10$), these two parameters exhibit an obvious increase with $l$ when $l$ is in a higher range (e.g., $l$ between 1 and 20). This phenomenon is often overlooked in experiments due to the challenges associated with characterizing rapidly varying spatiotemporal phases at large $l$ values. Fortunately, the high precision of our computational approach enables us to uncover this subtle yet significant behavior. Additionally, the radius $r$ and the thinness of these STOVs are also influenced by the conical phases ($\alpha r$) used in their generation. Based on these observations, we classify the generated wavepacket as a QPSTOV rather than a PSTOV.

The generated QPSTOVs with different TCs using $\alpha$ = 8662 rad/m are illustrated in Fig.2. Rows 1 and 2 show the spatiotemporal intensity profiles and phases of the QPSTOVs in the near field, respectively. Row 3 presents the three-dimensional (3D) iso-intensity profiles of the wave packets of the generated QPSTOVs in the far field, viewed in the $t$-$y$ plane. Rows 4 and 5 show the spatiotemporal intensity profiles and phases of the QPSTOVs in the far field, respectively. The TCs of the generated QPSTOVs are 1, 5, 10, 15, and 20. For comparison, a QPSTOV with $l = 0$ is shown in column 1.

In the near field, the QPSTOVs exhibit a ring-shaped profile in the time ($t$) direction and a Bessel-shaped profile in the space ($y$) direction. The phase jumps of the QPSTOVs along the $y$-direction are clearly visible in row 2. In the far field, ring-shaped STOVs are observed. For QPSTOVs with low $l$, such as $l < 10$, the radius $r$ and the thinness of the QPSTOV remain nearly constant, However, significant deviations in these parameters become apparent as the TC increases. This indicates that the energy of the QPSTOV cannot be effectively confined within the same thin ring as the TC grows, a behavior that distinguishes it from spatial perfect vortices. In the case of spatial perfect vortices with varying TCs, the energy remains well localized within nearly the same thin ring [34].

For QPSTOV with $l = n$, the phase patterns exhibit $n \times 2\pi$ phase winding. Notably, the spiral phase of QPSTOV with large TCs, such as $l = 20$, is also well preserved. In contrast, the QPSTOV with $l = 0$, lacks a spiral phase in its phase pattern and instead manifests as a spatiotemporal Bessel beam [36].

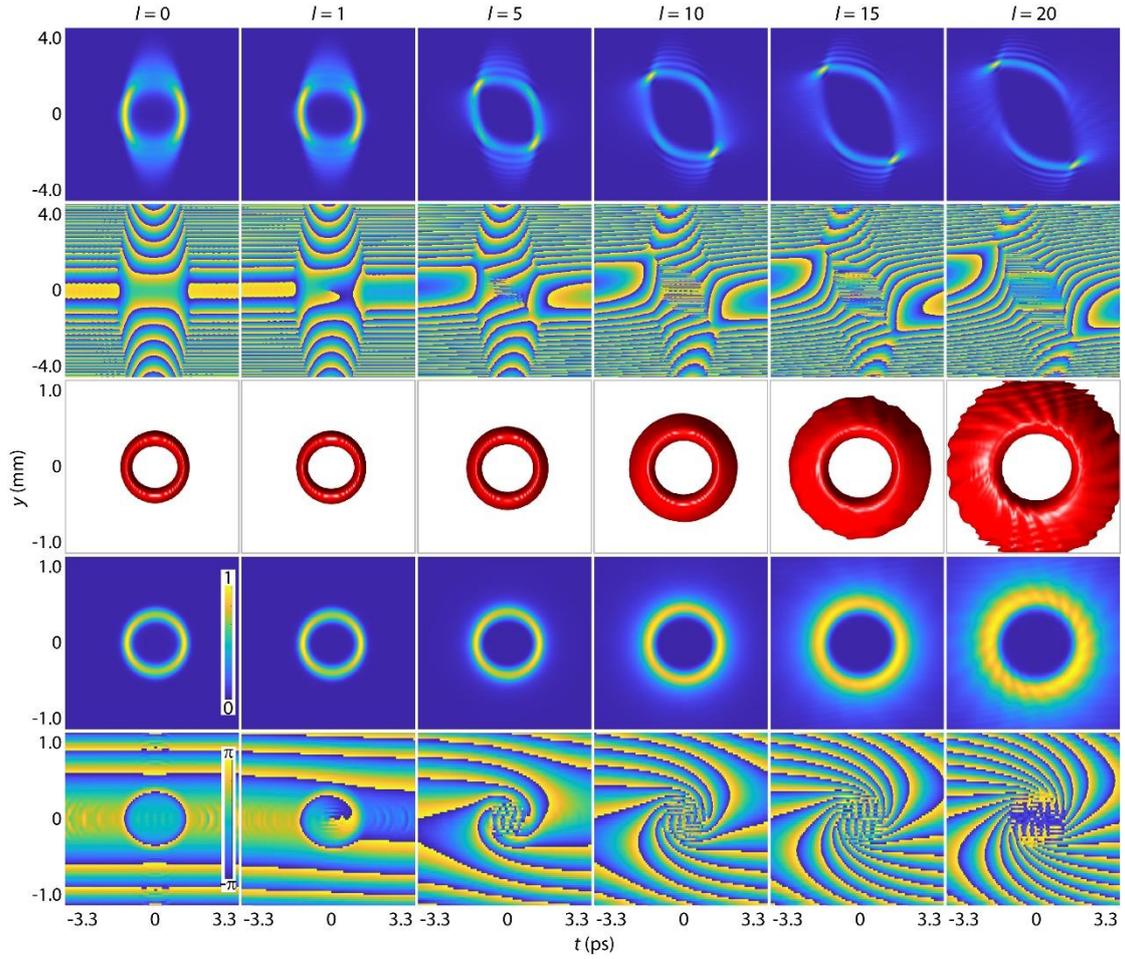

Fig. 2. The calculated intensity profiles and phase patterns of QPSTOVs. Rows 1 and 2 show the intensity profiles and the corresponding phase patterns of the QPSTOVs in the near field, respectively. Row 3 shows the isosurface plots of the 3D (*x-y-t*) wave packets of the QPSTOVs generated in the far field, and they are viewed in the *t-y* plane. Rows 4 and 5 show the intensity profiles and the corresponding phase patterns of the QPSTOVs in the far field, respectively. The top row is marked as row 1.

The diameters of the QPSTOVs generated using three different conical phases in the spatial and temporal directions are demonstrated in Fig. 3, the three corresponding axicon parameters are $\alpha_1 = 4331$ rad/m, $\alpha_2 = 8662$ rad/m, and $\alpha_3 = 17324$ rad/m. For a given TC, the radii of the QPSTOV in both the *y*- and *t*- directions increase as the conical phases increase, which is equivalent to an increase in the base angle of the axicon lens if a physical axicon were employed. Moreover, the rate of increase in the radius of the QPSTOV slows as the conical phases become larger. The diameters of the QPSTOV are determined by measuring the distance between the two intensity peaks in both the *y* and *t* directions.

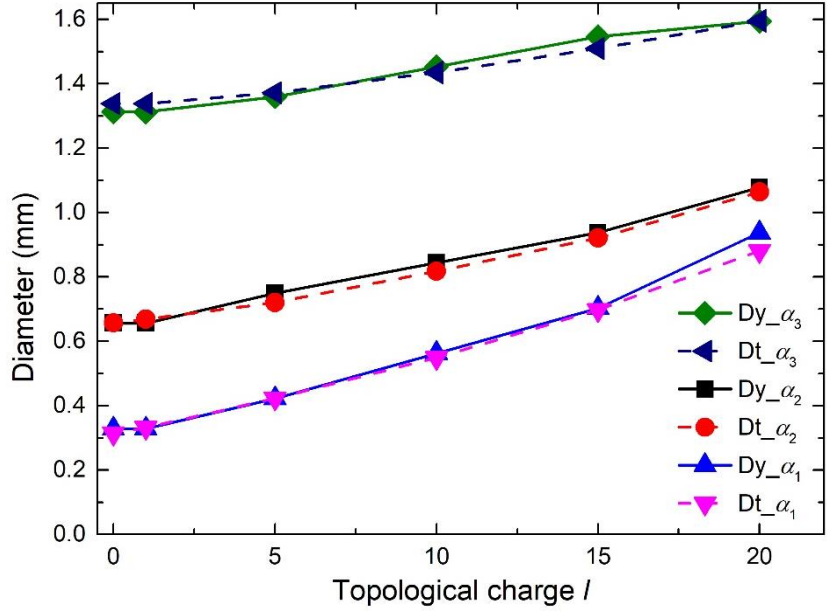

Fig. 3. The spatial and temporal diameters of the QPSTOVs with different TCs. The QPSTOVs are generated using three different conical phases, which are produced using three different axicon parameters, $\alpha_1$, $\alpha_2$, and $\alpha_3$, specifically $\alpha_3 = 4\alpha_1$ and $\alpha_2 = 2\alpha_1$.

The evolutions of QPSTOVs generated using spherical lens with different focal lengths in free space are exhibited in Fig. 4. For comparison, an elliptical QPSTOV generated using a spherical lens with a longer focal length ($f = 1$ m) is also shown in Fig. 4, as the generated QPSTOVs somewhile exhibit an elliptical shape. Row 1 of Fig. 4a shows the propagation of the QPSTOVs generated using a spherical lens with $f = 300$ mm. In the near field, the QPSTOV displays a ring-shaped profile in the time ($t$) direction and a Bessel-shaped profile in the space ($y$) direction, as previously discussed. During propagation, the $y$-radius of the QPSTOV decreases while the $t$-radius remains constant, resulting in a circular QPSTOV at the focal plane. Beyond the focal plane, the QPSTOV becomes elliptical and tilts in the opposite direction relative to its pre-focal plane orientation. The QPSTOV collapses at $z = 1.2f$ and forms a pattern resembling an interference structure. The evolution of the corresponding phase is shown in row 1 of Fig. 4a. The QPSTOV exhibits an asymmetric evolution.

For the elliptical QPSTOV generated using the $f = 1$ m lens, the focusing effect in the $y$-direction is mitigated, and the collapse occurs at a greater distance beyond the focal plane. Notably, the $t$-radii of both QPSTOVs remains unchanged during the propagation, as no dispersion is introduced in the process. Interestingly, within distances shorter than the focal length, the QPSTOV does not exhibit the temporal diffraction effect that typically causes mode degradation in STOVs during propagation, such as splitting a high-order STOV into a wave packet composed of multiple first-order STOVs [18]. Additionally, no Bessel STOV is observed during the transition from the near field to the far field.

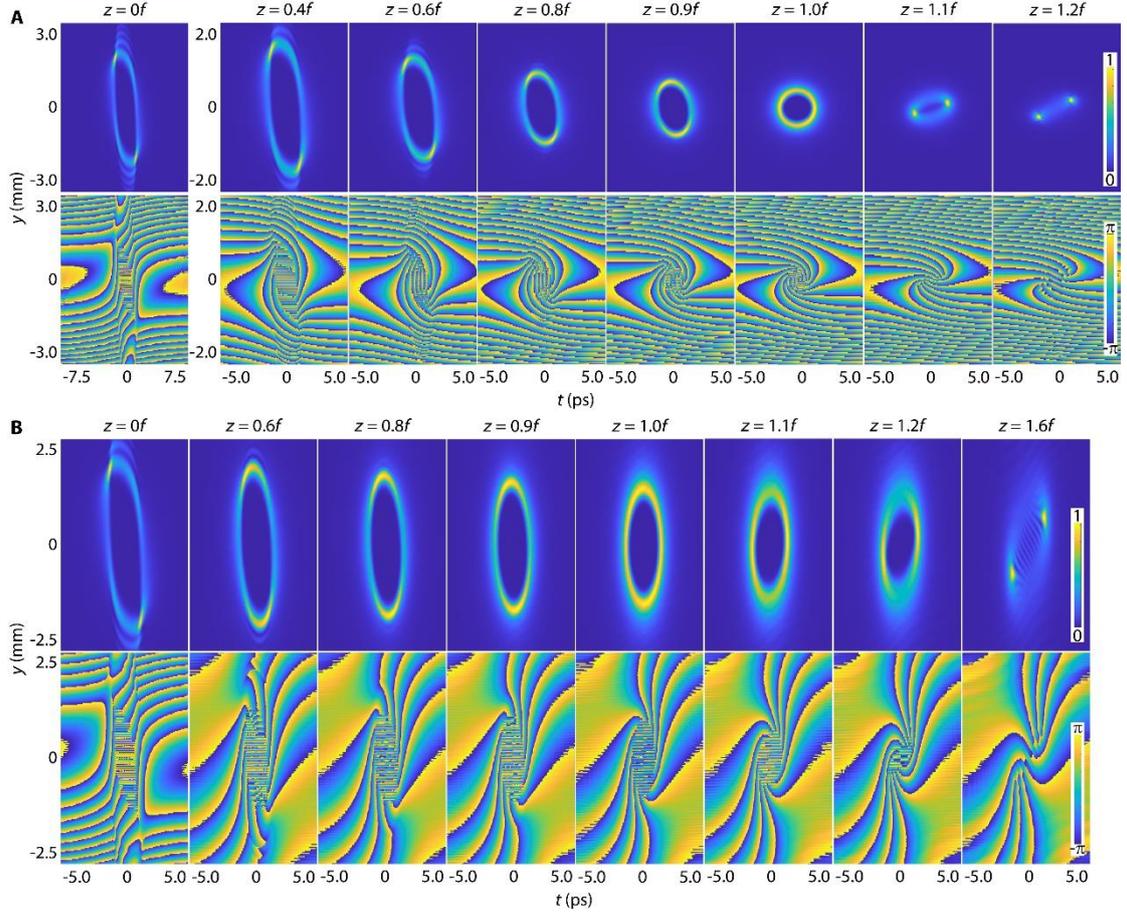

Fig. 4. The evolution of QPSTOVs in free space. (a) shows the intensity profiles (top row) and the corresponding phase (bottom row) of the QPSTOV generated using $f$ = 300 mm spherical lens at different locations during propagation. (b) shows the intensity profiles (top row) and the corresponding phase (bottom row) of the QPSTOV generated using $f$ = 1000 mm spherical lens at different locations during propagation.

The effect of group delay dispersion (GDD) on the generated QPSTOVs is illustrated in Fig. 5. The GDD values indicated at the top of each column represent the dispersion added to the input laser pulses, where $GDD_0$ = 91869 $fs^2$. In the near field, the wave packet generated with GDD = 0 transitions from a $y$-Bessel and $t$-ring shape to a Bessel-shaped wave packet in both the $y$- and $t$- directions as the GDD increases, while it collapses as the GDD decreases. In the far field, the generated QPSTOV evolves into an ellipse shape with an increase $t$-radius, and forms a $t$-Bessel and $y$-ring shape at GDD = 1.5$GDD_0$. The $t$-radius decrease with the decrease of GDD, leading to the collapse of the wave packet. Furthermore, the *y*-radius is maintained when the GDD is change. The corresponding phases of the wavepackets are shown in row 4.

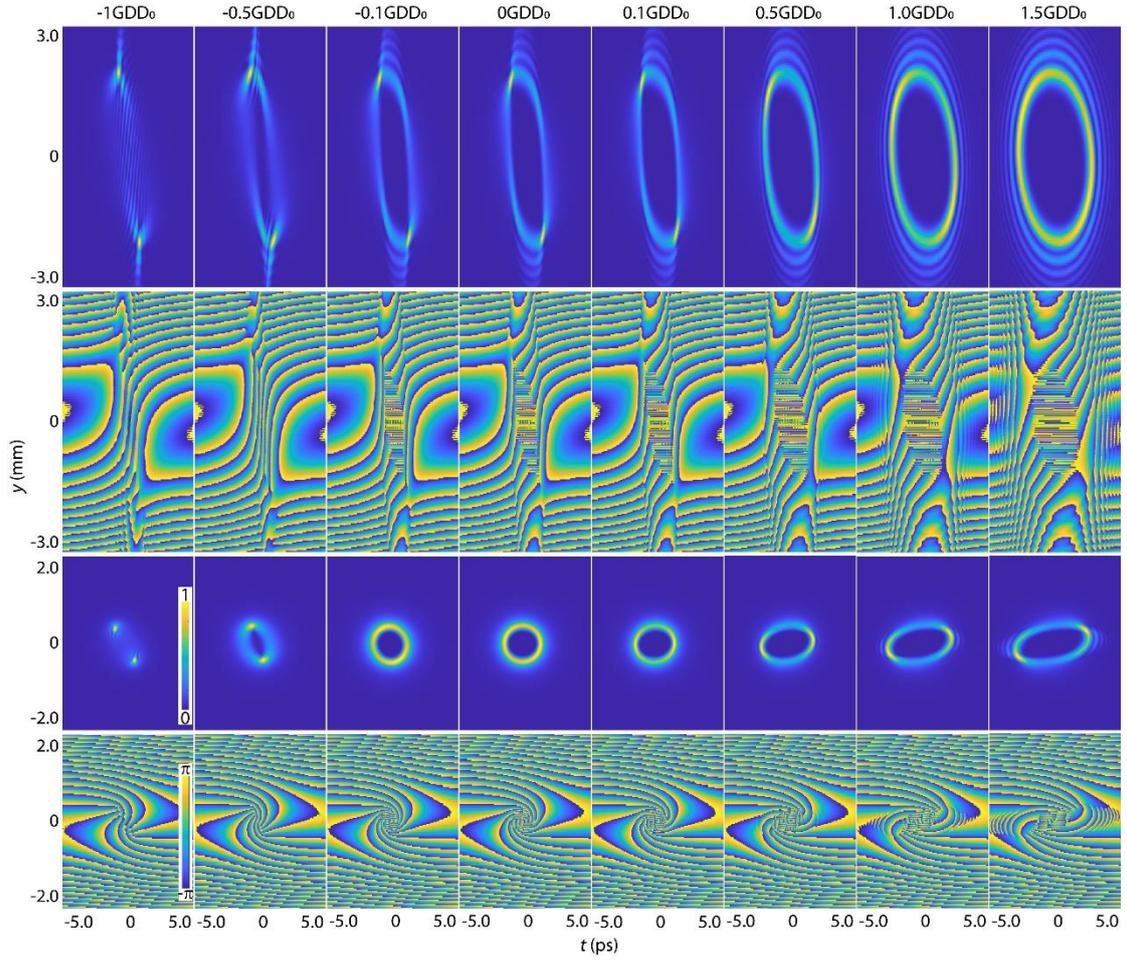

Fig. 5. The intensity profiles and phase patterns of the QPSTOVs generated with different GDD. (a) shows the intensity profiles (top row) and the corresponding phases (bottom row) of the QPSTOVs generated with different GDD in the near field. (b) shows the intensity profiles (top row) and the corresponding phases (bottom row) of the QPSTOVs generated with different GDD in the far field.

From Figs. 4 and 5, it is evident that the *y*-radius and *t*-radius of the QPSTOVs can be adjusted by controlling the focal length of the spherical lens and the GDD of the input laser pulses, respectively. Notably, a ring-shaped QPSTOV with a uniform intensity distribution can be achieved within a GDD range of $\pm 0.1 GDD_0$, which significantly relaxes the tolerance requirements for the GDD of the input pulse during QPSTOV generation. Additionally, a GDD of $0.2 GDD_0$ is considerable for a laser pulse centered at 800 nm if accumulated during propagation in air, as it would require a propagation distance of approximately 1 km to achieve such dispersion. This implies that the QPSTOV is highly robust during propagation, and its ring-shaped profile can be maintained over long distances, up to about 1 km. By using laser pulses centered at longer wavelengths, which experience lower GDD in air, the QPSTOV can propagate even farther.

## 4. Conclusion

The QPSTOV (or MDS-STOVs) can be generated by performing a 2D STFT of a Bessel beam in the $y$-$\omega$ domain. The radius and thinness of the QPSTOV are not entirely independent of the topological charge but instead exhibit a moderate dependence on it. This observation highlights a fundamental distinction between generating perfect vortices in the traditional spatial domain and in the spatiotemporal domain. Furthermore, the evolution of QPSTOVs and the impact of GDD have been thoroughly analyzed, revealing the robustness in both generation and transmission. Additionally, a mode-degradation suppression or time-diffraction-free property is observed as the QPSTOV propagates before passing through the focal plane, further underscoring its unique characteristics. This work enhances our understanding of the physical properties of SOTVs, and will promote the application of STOV in optical communication, quantum information, light-matter interaction, etc. The insights gained from this research are expected to drive further advancements in these areas, leveraging the unique properties of STOVs for innovative applications.

**Funding.** National Natural Science Foundation of China (NSFC) (12374320); Natural Science Foundation of Shanghai (23ZR1471700).

**Disclosures.** The authors declare no conflicts of interest.

**Data availability.** Data underlying the results presented in this paper are not publicly available at this time but may be obtained from the authors upon reasonable request.